\documentclass[preprint,preprintnumbers,amsmath,amssymb]{revtex4}
\usepackage{graphicx}
\begin{document}

%
%
\title{Multi pulse control of entanglement}  
\author{Chikako Uchiyama\(^{1}\) and Masaki Aihara\(^{2}\)}
\affiliation{%
\(^{1}\)Interdisciplinary Graduate School of Medicine and Engineering, 
University of Yamanashi,\\
4-3-11, Takeda, Kofu, Yamanashi 400-8511, JAPAN,\\
\(^{2}\)Graduate School of Materials Science,
Nara Institute of Science and Technology,\\
8916-5, Takayama-cho, Ikoma, Nara 630-0101 JAPAN
}%
\date{\today} 
%
\def\ch{{\cal H}}
\def\hbar{\mathchar'26\mkern -9muh}
\def\muv{{\vec \mu}}
\def\ev{{\vec E}}
\def\ih{\frac{i}{\hbar}}
\def\taus{{\tilde \tau}_{s}}
\def\tauc{{\tilde \tau}_{c}}
\def\ee{|e\rangle \langle e|}
\def\eg{|e\rangle \langle g|}
\def\ge{|g\rangle \langle e|}
\def\gg{|g\rangle \langle g|}

\begin{abstract}
We study the effectiveness of multi pulse control to suppress the degradation of entanglement.  Based on a linearly interacting spin-boson model, we show that the multi pulse application recovers the decay of concurrence when an entangled pair of spins interacts with a reservoir that has the non-Markovian nature. We present the effectiveness of multi pulse control for both the common bath case and the individual bath case.  
\end{abstract}

\maketitle

\section{Introduction}
\label{sec:1}
Quantum entanglement plays a central role in quantum information processing such as quantum teleportation\cite{bennett93}, and quantum computation\cite{chuang00}.  However, the purity of entanglement is vulnerable to various environmental effects, which is an obstacle to realize these quantum information processing.  

In this paper, we show that the multi pulse application can suppress the degradation of entanglement for a pair of qubits by focusing on concurrence as the degree of entanglement. We consider that the non-Markovian nature of the reservoir is the key to suppress the decay of purity of entanglement.  The method does not require high accuracy measurement, classical communication, or decreasing the number of pairs which have been requisite to execute entanglement concentration, purification or distillation\cite{bennett96,deutsch96,bennett962,bennett961,duan00,bose99,zeilinger1,zeilinger2,kwiat,yamamoto} .

\section{Preliminaries}
\label{sec:2}
Concurrence, a measure of purity of entanglement, has been introduced by Wootters\cite{wooters97,wooters98,wooters01} as 
\begin{equation}
C(\rho) = {\tt max} (0, 2{\lambda_{{\tt max}}} - {\tt Tr} R).
\label{eqn:1}
\end{equation} 
Here \(\lambda_{{\tt max}}\) is the maximum eigenvalue of the operator \(R\)  which is defined by
\begin{equation}
R = \sqrt {\sqrt{\rho} {\tilde \rho} \sqrt{\rho}},
\label{eqn:2}
\end{equation} 
for the density matrix of the pair of qubits \(\rho\) and \({\tilde \rho}\),
\begin{equation}
{\tilde \rho} = (\sigma_{1,y} \otimes \sigma_{2,y}) \rho^{*} (\sigma_{1,y} \otimes \sigma_{2,y}),  
\label{eqn:3}
\end{equation} 
where the \(y\)-component of the Pauli matrix, \(\sigma_{n,y}\), is associated with the spin-flip operation on the \(n\)-th qubits. In Eq.(\ref{eqn:3}), \(\rho^{*}\) is the complex conjugate of \(\rho\).  Since the complex conjugation is associated with time reversal of \(\rho\)\cite{terhal03}, \({\tilde \rho}\) comprises spin-flip and time reversal operations on \(\rho\). 

The quantity of \({\tt Tr} R\) indicates the degree of equality between\(\rho\) and \({\tilde \rho}\) \cite{bures,uhlman,josza,wooters97,wooters98,wooters01}.  This means that the concurrence reflects the ``degree of equality" between a density matrix under consideration and a density matrix obtained by spin-flip and time reversal.  By this definition, we can consider that the concurrence describes the degree of reversibility of a pair of qubits after spin-flip, which suggests that, if we can control the time reversibility, we can directly control the entanglement.   

\begin{figure}[h]
\begin{center}
\includegraphics[scale=0.4]{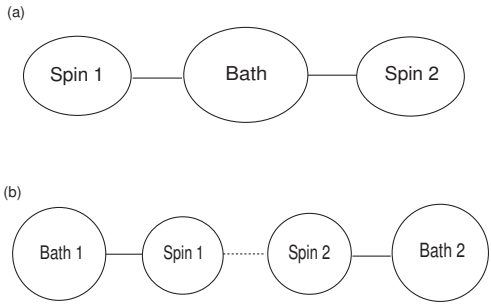}
\end{center}
\caption{Scheme of models: (a) common bath case and (b) individual bath case.} 
\label{fig:fg1}
\end{figure}

Control of time reversibility on a single qubit have been studied in our previous papers\cite{uchiyama1,uchiyama2} by providing the physical background of the suppression of decoherence with multi pulse application, which is categorized to the bang-bang method\cite{lloyd1,ban,luming,lloyda,lloydb,viola2,viola3,henryk,lidars}.  We have shown that the non-Markovian nature of a reservoir plays an essential role to determine the degree of the effectiveness of suppression of decoherence of a single qubit. When a time evolution of a reservoir has non-Markovian nature, or memory effect, the decoherence of a qubit is partial reversible.  Since \(\pi\) pulse application causes a time reversal operation to a qubit, the time evolution of the qubit is partially reversed.  We extend this to the case of a pair of qubits by using a linearly interacting spin-boson model.   Let us consider two extreme cases schematically shown in Fig.\ref{fig:fg1}: 1) common bath case: a pair of qubits interact with a common reservoir, 2) individual bath case: each qubit interacts with its own bath.

\section{Formulation}
\label{sec:3}

\subsection{Common bath case}
\label{sec:31}
When a pair of \(\frac{1}{2}\) spins linearly interact with a common reservoir, the Hamiltonian of this system is given by
\begin{equation}
\ch_{R}= \ch_{S} + \ch_{B} + \ch_{SB}\;,
\label{eqn:4}
\end{equation}
where
\begin{eqnarray}
\ch_{S}&=& \hbar \sum_{n=1}^{2} \omega_{0} S_{n,z} \label{eqn:5}\\
\ch_{B}&=&\sum_{k} \hbar \omega_{k} b_{k}^{\dagger} b_{k} \label{eqn:6}\\
\ch_{SB} &=& \hbar \sum_{n=1}^{2} \omega S_{n,z}  \sum_{k} h_{k} \omega_{k} (b_{k}+ b_{k}^{\dagger}) \label{eqn:7}\; . 
\end{eqnarray} 
Here  \(S_{n,z} \) is the z-component of the \(n\)-th \(\frac{1}{2}\) spin \((n=1,2)\), \( b_{k} (b_{k}^{\dagger}) \) is the annihilation (creation) operator of \(k\)-th boson which composes the bath, and \(h_{k}\) is the coupling strength between the spin and the \(k\)-th boson of the bath.  This model is applicable to a pair of quantum dots in semiconductors\cite{braun}. 

When an entangled qubit pair interacts with a noisy environment, the concurrence decays from the maximum value to zero, which indicates that the pair loses purity.  To suppress the decay of purity, let us consider the application of sufficiently short \(\pi\) pulse sequence on both spins. The Hamiltonian under pulse application is written as
\begin{eqnarray}
\ch_{SP}(t) &=& \ch_{S} + \sum_{j=1}^{N} \ch_{P,j}(t), \label{eqn:8} \\
\ch_{P,j}(t) &=& -\frac{1}{2} \ev_{j}(t) \cdot \muv \; \sum_{n=1}^{2} (S_{n,+} e^{-i \omega_{0} t} +S_{n,-} e^{i \omega_{0} t})  \label{eqn:9} 
\end{eqnarray} 
where \(\ev_{j}(t)\) indicates the field amplitude of \(j\)-th applied pulse.  In Eq. (\ref{eqn:9}), we assume that the both spins have the same magnetic moment \(\muv \) and we apply the pulse field which is on resonance with the magnetic moment. When we read a 1/2 spin as a two-level system, we can apply Eq.(\ref{eqn:4}) to the electric interaction for optical transition as well as to the magnetic interaction.  

Let us focus on the time evolution of a pair of spins which are maximally entangled at an initial time as \(|\psi\rangle_{t=0} =\frac{1}{2}(|1\rangle|1\rangle +|0\rangle|0\rangle)\). We assume the boson bath to be in the vacuum state at an initial time and an coupling function to be a Gaussian distribution with the mean frequency \(\omega_{p}\) and the variance \(\gamma_{p}\) as,
\begin{equation}
h(\omega) \equiv \sum_{k} |h_{k}|^2 \delta(\omega-\omega_k) 
\equiv \frac{s }{\sqrt{\pi} \gamma_{p} } \exp(-\frac{(\omega-\omega_{p})^2}{\gamma_{p}^2}),\label{eqn:10} 
\end{equation} 
where \(s\) means the average number of bosons interacting with a spin.
We obtain the time evolution of concurrence under \(N\) times \(\pi\) pulse application with pulse interval \(\tau_{s}\) as,
\begin{equation}
C(t)= \frac{1}{2} \exp[-2 \sum_{k} |\alpha _{k} (t)|^2]
\label{eqn:11}
\end{equation} 
where 
\begin{eqnarray}
\alpha _{k} (t) &=& h_k e^{ - i\epsilon _k (t - N\tau _s )} \{( {1 - e^{i\epsilon _k (t - N\tau _s )} }) \nonumber \\
&& + \sum_{m = 1}^{N} {(- 1)^m e^{ - i m \epsilon _k \tau _s } } (1 - e^{ - i\epsilon _k \tau _s } ) \}. \label{eqn:12}
\end{eqnarray} 

\subsection{Individual bath case}
\label{sec:32}
Next, we consider the individual bath case where each spin individually interacts with its own boson reservoir as 
\begin{equation}
\ch_{SB} = \hbar \sum_{n=1}^{2} \omega S_{n,z}  \sum_{k_{n}} h_{k_{n}} \omega_{k_{n}} (b_{k_{n}}^{(n)} + {b_{k_{n}}^{(n)}}^{\dagger})\; , 
\label{eqn:13}
\end{equation} 
where \(b_{k_{n}}^{(n)}\)  \(({b_{k_{n}}^{(n)}}^{\dagger})\) is the creation (annihilation) operator of boson reservoir with which the \(n\)-th spin interacts.  In Eq.(\ref{eqn:13}),  \(\omega_{k_{n}}\) indicates the frequency of the \(k\)-th boson of the \(n\)-th bath. \(h_{k_{n}}\) is the coupling strength between the \(n\)-th spin and the \(k\)-th boson.  

Using the same initial condition for the spin state and boson bath as the common bath case, we obtain the time evolution of concurrence in the form as,
\begin{equation}
C(t)= \frac{1}{2} \prod_{n=1}^{2} \exp[- \frac{1}{2} \sum_{k_n} |\alpha _{k_n} (t)|^2],
\label{eqn:14}
\end{equation} 
where \(\alpha _{k_n} (t)\) is given by replacing suffixes \(k\) in Eq.(\ref{eqn:12}) with \(k_n\).  When we set the same coupling function for two baths, \(ln C(t)\) for individual bath case is a half of that in the common bath case in Eq.(\ref{eqn:11}).  This means that the qualitatively the same effect of a \(\pi\) pulse train is observed for the individual bath case.  

\section{Numerical evaluation}
\label{sec:4}
\begin{figure}[h]
\begin{center}
\includegraphics[scale=0.35]{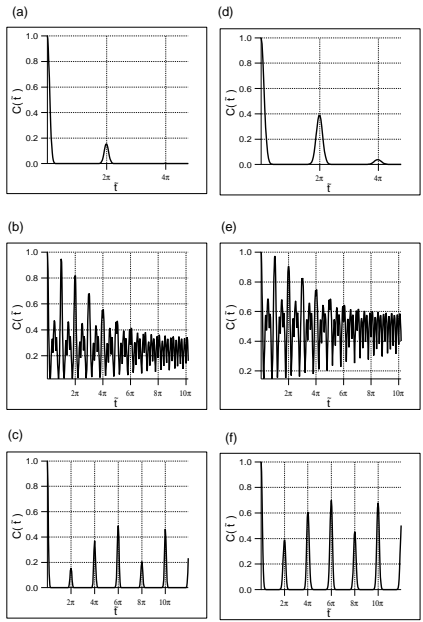}
\end{center}
\caption{Time evolution of \(C({\tilde t} )\) for \({\tilde \gamma_p \equiv \gamma_p/\omega_p} =0.1\), \( s= 5\). Figures (a),(b), and (c) correspond to the common bath case, and figures (d),(e), and (f) correspond to the individual bath case.  (a) and (d) show the case without pulse application; (b) and (e) for pulse interval \(\tilde{\tau}_{s}=\pi/5\); (c) and (f) for \(\tilde{\tau}_{s}=2 \pi\)}
\label{fig:fg2}
\end{figure}

Using a scaled time variable \( {\tilde t} \equiv t \omega_{p}\), we show the time dependence of the concurrence \(C({\tilde t})\) in Fig.\ref{fig:fg2}, where the parameters are set as \({\tilde \gamma_p\equiv \gamma_p/\omega_p}=0.1 \) and \( s= 5\).   Figs.\ref{fig:fg2}(a),\ref{fig:fg2}(b),and \ref{fig:fg2}(c) correspond to the common bath case, and Figs.\ref{fig:fg2}(d),\ref{fig:fg2}(e), and \ref{fig:fg2}(f) correspond to the individual bath case. Figures \ref{fig:fg2}(a) and \ref{fig:fg2}(d) show the case without pulse application, where we see a damped oscillation whose mean period is \( 2 \pi\) in the scaled time.  We can see that the decay of concurrence in the common bath case (Fig.\ref{fig:fg2}(a)) is faster than that in the individual bath case(Fig.\ref{fig:fg2}(d)).  This arises from the difference of coefficients in the exponents of Eq.(\ref{eqn:11}) and Eq.(\ref{eqn:14}).
Next, we show the effects of a \(\pi\) pulse train with a relatively short interval \(\tilde{\tau}_{s}=\pi/5 \) in Figs. \ref{fig:fg2}(b) and \ref{fig:fg2}(e). One can see that the decay of concurrence shows oscillation and approaches to a constant value.  This result clearly indicates that the degradation of concurrence is effectively suppressed by the application of \(\pi\) pulse train for both the common and individual bath case. While the degree of suppression decreases as increasing the pulse interval to \(\tilde{\tau}_{s}=\pi\), the concurrence periodically recovers by synchronizing the pulse interval to the oscillation period corresponding to the center frequency of the coupling function by setting \(\tilde{\tau}_{s}=2 \pi\) as shown in Figs. \ref{fig:fg2}(c) for common bath case and \ref{fig:fg2}(f) for individual bath case.  

\section{Conclusion}
\label{sec:4}
Focusing on the time evolution of concurrence, we have shown that the multi pulse application can suppress the degradation of the quantum entanglement in the individual bath case as well as the common bath case.  We found that the non-Markovian nature of bath plays an essential role to determine the effectiveness of the multi pulse control.    

While we have used the concurrence as a measure of degree of entanglement in this paper, we can evaluate the time evolution of entropy and purity with the concurrence for this model\cite{uchiyama3}.

\end{document}